\documentclass[sigconf]{acmart}

\usepackage{graphicx} 
\usepackage{amsmath}
\usepackage{enumitem}
\newtheorem{myDef}{Definition}
\usepackage{booktabs}
\usepackage{multirow}
\usepackage{makecell}

\usepackage[most]{tcolorbox} 
\usepackage{listings}        
\tcbuselibrary{breakable}

\AtBeginDocument{%
  }

\copyrightyear{2026}
\acmYear{2026}
\setcopyright{cc}
\setcctype{by-nc-nd}
\acmConference[WWW '26]{Proceedings of the ACM Web Conference 2026}{April 13--17, 2026}{Dubai, United Arab Emirates}
\acmBooktitle{Proceedings of the ACM Web Conference 2026 (WWW '26), April 13--17, 2026, Dubai, United Arab Emirates}
\acmPrice{}
\acmDOI{10.1145/3774904.3792813}
\acmISBN{979-8-4007-2307-0/2026/04}

\begin{document}
\title{Data-Driven Function Calling Improvements in Large Language Model for Online Financial QA}

\author{Xing Tang}
\authornote{Both authors contributed equally to this research.}
\affiliation{%
  \institution{Shenzhen Technology University}
  \city{Shenzhen}
  \country{China}
}

\author{Hao Chen}
\authornotemark[1]
\affiliation{%
  \institution{FiT, Tencent}
  \city{Shenzhen}
  \country{China}
}

\author{Shiwei Li}
\affiliation{%
  \institution{Huazhong University of Science and Technology}
  \city{Wuhan}
  \country{China}
}

\author{Fuyuan Lyu}
\affiliation{
  \institution{McGill University}
  \city{Montreal}
  \country{Canada}
}

\author{Weijie Shi}
\affiliation{
  \institution{The Hong Kong University of Science and Technology}
  \city{Hong Kong SAR}
  \country{China}
}
\author{Lingjie Li}
\affiliation{%
  \institution{Shenzhen Technology University}
  \city{Shenzhen}
  \country{China}
}
\author{Dugang Liu}
\affiliation{
  \institution{Shenzhen University}
  \city{Shenzhen}
  \country{China}
}

\author{Weihong Luo}
\authornote{Corresponding authors.}
\author{Xiku Du}
\affiliation{%
  \institution{FiT, Tencent}
  \city{Shenzhen}
  \country{China}
}

\author{Xiuqiang He}
\authornotemark[2]
\affiliation{
  \institution{Shenzhen Technology University}
  \city{Shenzhen}
  \country{China}
}

\renewcommand{\shortauthors}{Xing Tang, et al.}
\begin{abstract}
Large language models (LLMs) have been incorporated into numerous industrial applications. Meanwhile, a vast array of API assets is scattered across various functions in the financial domain. An online financial question-answering system can leverage both LLMs and private APIs to provide timely financial analysis and information. The key is equipping the LLM model with function calling capability tailored to a financial scenario. However, a generic LLM requires customized financial APIs to call and struggles to adapt to the financial domain.
Additionally, online user queries are diverse and contain out-of-distribution parameters compared with the required function input parameters, which makes it more difficult for a generic LLM to serve online users. In this paper, we propose a data-driven pipeline to enhance function calling in LLM for our online, deployed financial QA, comprising dataset construction, data augmentation, and model training. Specifically, we construct a dataset based on a previous study and update it periodically, incorporating queries and an augmentation method named AugFC. The addition of user query-related samples will \textit{exploit} our financial toolset in a data-driven manner, and AugFC explores the possible parameter values to enhance the diversity of our updated dataset.
Then, we train an LLM with a two-step method, which enables the use of our financial functions. Extensive experiments on existing offline datasets, as well as the deployment of an online scenario, illustrate the superiority of our pipeline. The related pipeline has been adopted in the financial QA of YuanBao\footnote{https://yuanbao.tencent.com/chat/}, one of the largest chat platforms in China.

\end{abstract}


  
\maketitle

\section{Introduction}
\label{sec:intro}
Recently, large language models (LLMs) have emerged as a powerful tool, demonstrating remarkable capabilities in understanding, generating, and reasoning with text~\cite{deepseek,gpt,qwen}. These features enable LLMs to seamlessly integrate with various web applications, including online code copilots, online chatbots, and question-answering systems. In finance and economics, various financial documents are used to analyze and predict market trends~\cite{FinBert}. Therefore, equipped with LLMs, online financial question-answering (QA) systems have shown promising progress in understanding and responding to complex queries related to these financial documents~\cite{survey,finben}.

\begin{figure*}
    \centering
    \includegraphics[width=0.85\linewidth]{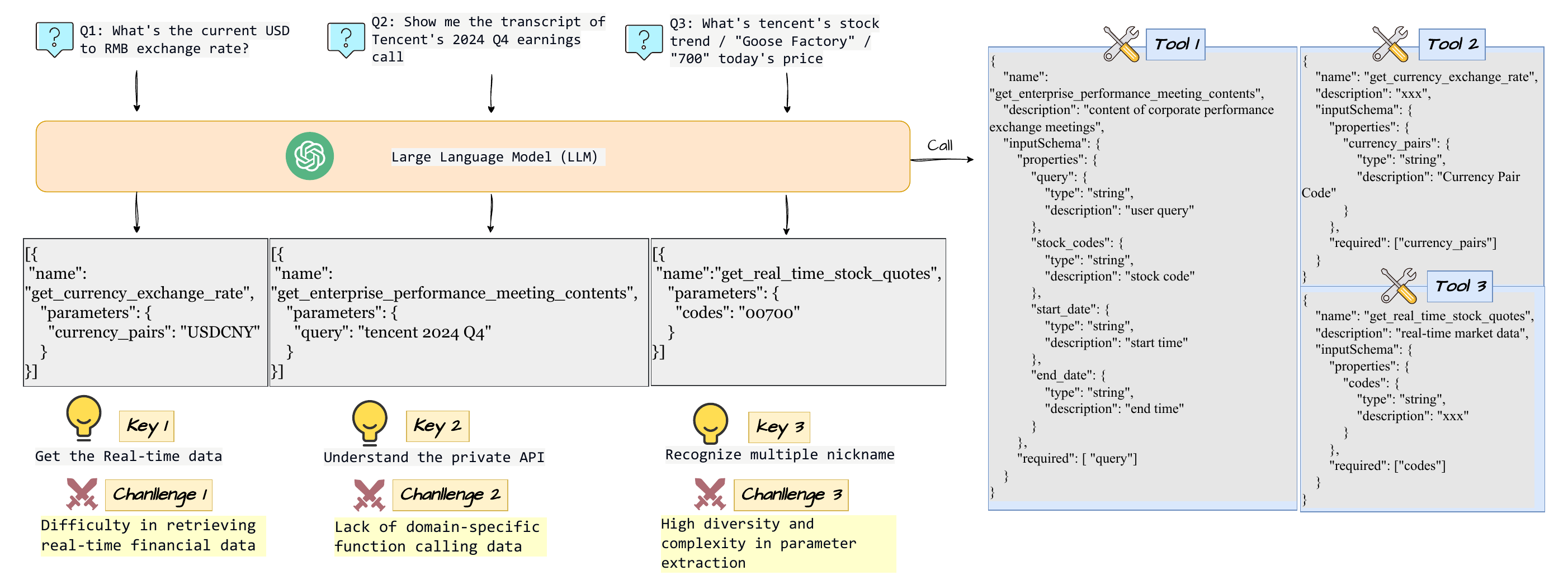}
    \caption{The overview of the key points and corresponding challenges in the online financial QA system powered by LLM.}
    \label{fig:intro}
\end{figure*}

Building an online financial QA system powered by LLM is non-trivial and requires specific efforts. Typically, in the financial scenario, important financial information is often provided by external APIs or functions and must be updated promptly, which restricts the direct application of LLM in financial QA, as illustrated in the first key point of Figure.~\ref {fig:intro}. To address this, the LLM can be trained to utilize timely external knowledge via function calling or tool calling~\cite{fcsurvey,APIbank,APIgen,Hammer}, a technique that has been widely adopted in agents~\cite{agent,react}. The core components of function calling are, respectively, tool selection and parameter extraction based on the query and function documents. Provided private APIs and functions, LLM can retrieve the external up-to-date knowledge based on the extracted parameters from the user query. Together with its internal capability, the empowered LLM can give a more accurate answer.

A simple approach is to introduce a commercial LLM equipped with general function calling capabilities, directly serving user queries~\cite{kimi}.
While commercial LLMs generally perform well in function calling, these models often struggle to provide accurate and robust function calling capabilities for specialized scenarios due to the lack of private training data~\cite{ESFC}. As the example in Figure.~\ref{fig:intro} indicates, the function used to 
"get enterprise performance meeting contents" is usually unique to financial analysis, which is unavailable in the general function calling dataset. Hence, collecting function calling data to train a financial tool-specific LLM is essential.
Moreover, the financial APIs are usually highly customized, while the queries are diverse. The third key point in Figure.~\ref {fig:intro} illustrates this challenge. Invoking the function "get stock quotes" requires extracting the company name as a parameter from the user query. Some queries will directly specify the company name, while some will give a nickname. For example, both "700" and "Goose Factory" in queries refer to Tencent, where the code "700" is the stock symbol of Tencent, and "Goose Factory" is the mascot of this company. Therefore, the diversity of private datasets based on the user queries poses another challenge in improving the performance of function calling. 

In this work, we design a data-driven pipeline to improve the function calling in LLM for our online financial QA. Starting with an annotated financial QA dataset following xLAM~\cite{xLAM}, our dataset is periodically updated with both user queries and augmented datasets. Specifically, constructing user query-related samples \textit{exploits} the existing toolset based on direct function call results in online interaction. This exploitation is responsible for improving coverage on the financial tool sets of our dataset. However, as previously stated, the diversity poses a challenge for both the query and the parameters. We thus propose an automated augmentation method named \textbf{AugFC} to \textit{explore} the possible queries containing parameter values in our datasets. Based on the updated dataset, we further train a language model that includes a supervised fine-tuning (SFT) stage and a reinforcement learning stage, enabling the base model to utilize financial tools aligned with our scenario.
In summary, our main contributions are as follows:
\begin{itemize}

\item We first identify the core challenges in building our online financial QA system, providing practical lessons from industrial applications.

\item We develop a data-driven automated function calling pipeline consisting of a dataset constructed, data augmentation, and model training to enhance the base LLM for our online financial QA system. The pipeline is effective in improving the performance of function calling in the LLM.

\item Extensive experiments on both the offline dataset and the online scenario have been conducted to validate the superiority of our method.
\end{itemize}
\section{Related work}

In this section, we provide a brief review of two topics related to our work: financial QA and function calling.

\subsection{Financial QA}
Question-answering systems have already achieved remarkable progress with the introduction of LLM. Most financial QA systems focus on numerical reasoning to handle 
multi-step calculations and extract relevant information from various data sources~\cite{finreasoning,finben,tat}. ZS-FinPYT and ZS-FinDSL~\cite{zshot,finfinetuning} introduce zero-shot techniques for LLMs to perform complex numerical reasoning over financial documents. A multi-agent framework is also adopted, incorporating a critic agent that reflects on the reasoning steps and final answers for each question~\cite{multi-agent}. Besides, some works are devoted to financial text QA~\cite{fintextqa}. WeaverBird~\cite{weaverbird} is a dialogue system specifically for the finance sector. Leveraging finetuned LLM on extensive financial corpora, it provides informed responses to complex user queries. Our work first gives how to incorporate function calling in LLM to solve diverse online financial QA queries, which hopefully sheds light on building industry financial QA.

\subsection{Function Calling}
Function calling or tool calling has represented a pivotal advancement in empowering LLMs with dynamic interaction capabilities in the external environment~\cite{react,toolalpaca}. The array of this field mainly focuses on two categories: data synthesis and model enhancement. There are plenty of data synthesis methods for constructing a general function calling dataset. Toolformer~\cite{toolformer} enhances the LLM's ability by finetuning the base model with API calling datasets. Then, ToolLLM~\cite{toolllm} collects 16,464 real-world APIs, including multi-tool usage, to finetune LLaMA and obtain ToolLLaMA. ToolACE~\cite{toolace} and xLAM~\cite{xLAM} utilize agents to collect tool use data, and also emphasize the validation process to filter data ~\cite{APIgen}. ToolHop~\cite{ToolHop} targets the multi-hop data with a query-driven data construction. Autotools~\cite{autotools} combines tool encapsulation and tool programming to empower LLM to automate the tool-use flow. All of these works focus on general function calling capabilities, ignoring how to build a dataset for a specific application, which is often abundant in interaction data.

The enhancing paradigm of the base model shifts from fine-tuning to reinforcement learning. Finetuning has been investigated based on the proposed datasets~\cite{toolace,toolllm,xLAM}. Some modifications are also proposed. Funreason~\cite{funreason} introduces a self-refinement multiscale loss to balance the reasoning and accuracy during finetuning. The enterprise-scenario function calling~\cite{ESFC} targets a specific domain and utilizes LoRa~\cite {lora} for finetuning. Reinforcement learning with verifiable reward has witnessed tremendous progress in LLM training~\cite{deepseek,kimi,funrl}.Tool-star~\cite{toolstar} and TooRL~\cite{toolrl} pioneer the application in tool calling. In our pipeline, we employ a two-step training paradigm and provide an LLM tailored for our financial QA.

\section{Methodology}
First, we will give a formal definition of our problem. Next, we will demonstrate our proposed data-driven pipeline, which includes data construction, data augmentation, and model training. We will elaborate on the design of each stage in this pipeline in detail. 

\subsection{Problem formulation}

For an online financial QA system powered by LLM denoted as $M$, there is a record of user queries $\mathbb{Q}$ and a toolset \footnote{We use the terms tool and function interchangeably as in a previous study.} $\mathbb{T}=\{t_1,t_2,\cdots,t_n\}$. For a particular user query $q\in \mathbb{Q}$, there is a corresponding reference tool-call list $a$ to solve the problem. For the toolset, the tool can be represented as $t_i=(name_i,description_i,parameters_i)$, where $name_i$ is the unique identifier of the tool, $description_i$ is the detailed functionality of the tool, and $parameters_i$ is the set of parameters used in this tool.
Let $\mathcal{P}$ denote the input prompt, which includes $q$ and $\mathbb{T}$, i.e. $\mathcal{P}=(q,\mathbb{T})$. Then the LLM will invoke the related tools, $a_g=M(\mathcal{P})$, where $a_g$ is the actual generated tool-call list.
Note that the $parameters_i$ define a set of required parameters. The LLM will determine the function $t_i$ and extract parameters $p_i$ from the query to invoke the related function.Then the generation can be further defined as, $a_g=[t_1(p_1),\cdots,t_m(p_m)]$ where $m$ is the total number of invoked functions.

Our goal is to construct the dataset $<q,a,\mathbb{T}>$ following xLAM format~\cite{xLAM}, and determine the model's policy $\pi:(q,\mathbb{T}) \to a$ with a set of rollouts $\{r_1,\cdots, r_L,t_1,\cdots,t_m\}$, where $r_1,\cdots,r_L$ are reasoning process.

\subsection{The Data-driven pipeline}
\begin{figure*}
    \centering
    \includegraphics[width=0.7\linewidth]{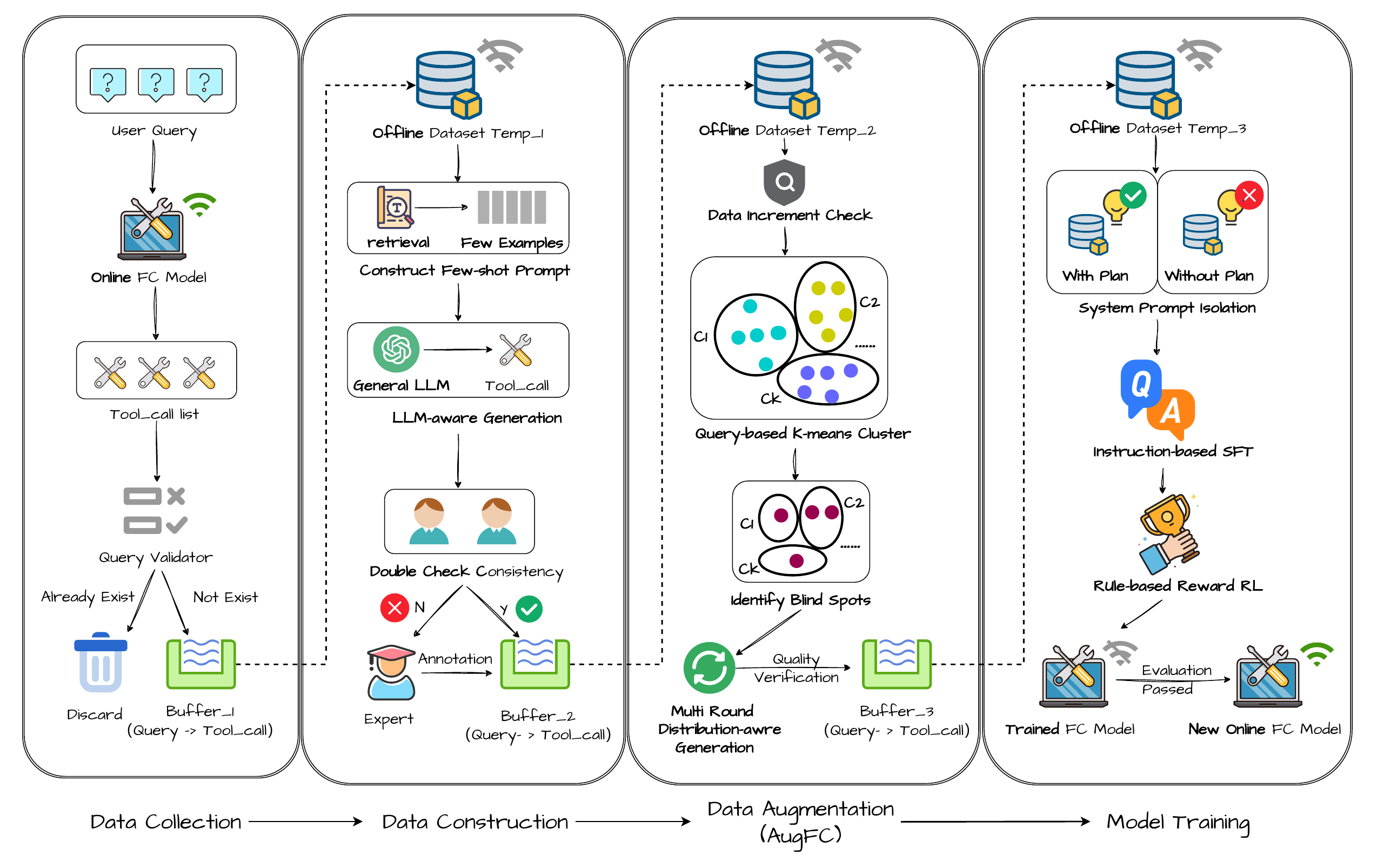}
    \caption{The data-driven pipeline consisting of data collection, data construction, data augmentation, and model training.}
    \label{fig:pipeline}
\end{figure*}

The overall data-driven pipeline is illustrated in Figure.~\ref{fig:pipeline}. In the initial setting, data construction will incorporate a small amount of manually annotated data as seed data, which provides a basis for the pipeline. There are four components in our pipeline. First, we will automatically collect online user queries. The data construction and augmentation will update the dataset in the updated stage. The first task is to update the datasets with online queries, which enables the dataset to exploit the user demands for the candidate function in our toolset. The second task involves exploring the diversity of queries, which utilize an automated method named \textbf{AugFC} to augment the query as necessary. Notice that online data fully drives both asks and updates the dataset aligned with the actual financial QA patterns.
Finally, the two-stage training is introduced to get the policy based on the updated dataset, considering effectiveness and efficiency. We will elaborate on these components in detail afterwards.

\subsubsection{Initial Settings}

Before we delve into the pipeline, we first introduce the seed data for the pipeline in our system. 

To ensure that the data closely aligns with our scenarios, we initially construct manually annotated data by financial experts. Notice that although the LLM can synthesize the data, the annotated data is better at reflecting the actual pattern in our financial QA system, and this will lay a good foundation for the following stage. We generally follow the xLAM format~\cite{xLAM}, which is denoted as the tuple$<q,a,\mathbb{T}>$. The experts will determine the $a$ from $\mathbb{T}$ for the selected query $q$. As stated in~\cite{ESFC}, the \textbf{diversity}, \textbf{uniqueness}, \textbf{consistency} are three principles in our annotation. The pipeline will be more stable and effective based on the high-quality seed data buffer $\mathbb{B}$.

\subsubsection{Data collection}

Although the annotated datasets cover a certain number of human-generated queries, users will likely provide more diverse queries in an online setting. Hence, utilizing these queries will enhance our dataset and reveal the actual pattern of how users make use of the QA system.

The first step in our pipeline is to collect the online queries. Once the user produces the query, the corresponding tool-call list is generated by LLM. The pair $<q,a_g,\mathbb{T}>$ will be set as a candidate for our data buffer $\mathbb{B}$. However, due to the number of queries being too large to handle, the query will undergo a validation process. With an embedding model $E$, the query can be validated if there have already been identical queries in our buffer. After validation, we denote the collected queries as $\mathbb{B}_g$, which should be constructed further and merged into $\mathbb{B}$.

\subsubsection{Data construction}

After getting the new candidate queries, we will construct the credible tool-call list for these queries. Notice that we already have a tool-call list $a_g$ from the online LLM. To preserve the expert's work, we introduce a more powerful LLM, $M_p$, to generate the tool-call list, which serves as a reference. Specifically, we retrieve the most similar queries in the buffer $\mathbb{B}$ to construct the few-shot prompting\footnote{All the prompt templates can be referred to the Appendix.}~\cite{few-shot}. The generated tool-call list $a_{M_g}$ by a powerful LLM will be compared with $a_g$ to double-check the consistency. The inconsistent queries between online LLM $M$ and powerful LLM $M_p$ will further be annotated by financial experts, and the consistent query will be merged into the buffer $\mathbb{B}$. 

In this process, we finally obtain high-quality question-tool pairs $<q,a_g,\mathbb{T}>$ by exploiting online user queries. Along with manually annotated pairs, we construct a dataset that aligns both financial experts' and online users' demands $\mathbb{B}=\mathbb{B}\cap\mathbb{B}_g$.

\subsubsection{Data augmentation} 

The online queries in our financial QA system are diverse, particularly in terms of parameter values, as stated in Section~\ref{sec:intro}. Typically, user-generated queries follow a power-law distribution~\cite{power} in parameter values, which renders our dataset inadequate to meet the diversity requirement in real-world scenarios.

To mitigate this issue, we propose an automatic data augmentation method, AugFC, to enhance the diversity in parameter values.
We first need to identify which parameter is the "blind spot", meaning the values of this parameter in our datasets have collapsed into a few single values. We introduce information entropy as a measure of the information contained in the set of parameter values.
Given the dataset buffer $\mathbb{B}$, the parameter value set is denoted as $p_j=\{p_j^i\}_{i=1}^N$ for each parameter $p_j$. The global entropy for $p_j$ can be calculated as follows,
\begin{equation}
H_{G}^{p_j}=-\Sigma_{p\in p_j}\frac{n_p}{N}\log_2\frac{n_p}{N},
\end{equation}
where $n_p$ is the count of the elements in the set. 

We then perform semantic clustering, which groups these queries into $K$ clusters based on the semantic embeddings of the queries. The tool will serve different semantic purposes for various semantic query clusters.
We also have a parameter value set $p_{j}^k=\{p_{j}^i\}_{i=1}^{N_k}$ in the cluster $k$.
The entropy for the $k$-th cluster can then be defined as follows,
\begin{equation}
H_{k}^{p_j}=-\Sigma_{p\in p_j}\frac{n_p^k}{N_k}\log_2\frac{n_p^k}{N_k}.
\end{equation}
We formally define the condition to determine whether one parameter is a blind spot.
\begin{myDef}
A parameter $p_j$ is called blind spot parameter $\iff$ $H_G^{p_j} > \tau_g$, and for each cluster $k$, $\frac{H_{k}^{p_j}}{H_G^{p_j}} < \tau_b$.
\label{def:blind}
\end{myDef}
The first condition indicates that the global entropy should exceed a threshold value. The reason is that certain global diversity needs to be guaranteed, and the parameter with smaller entropy should be exploited by updating new user queries during the data construction stage to ensure quality, rather than at this stage. The second one shows that the local diversity should not exceed a certain ratio compared to the global diversity, indicating that the distribution of the parameter collapses in this cluster.

With the identification of the blind spot of the parameter, we conduct multi-round distribution-aware generation, designing prompts for LLM $M_{aug}$ to generate the augmented data. Suppose the data can be denoted as $<q,t,p_b>$. We select the representative queries in cluster $k$, denoted as $\{q_k^{rep}\}$, as the context. Then the designed prompt contains the related information $\mathcal{P}_{aug}(\{q_k^{rep}\},q,H_G^{p_b},H_k^{p_b},\tau_b,\mathbb{T})$. The generated queries $q_{aug}=M_{aug}(\mathcal{P}_{aug})$ will update the dataset only if the cluster diversity is improved. Notice that the AugFC is fully automatic, requiring no manual intervention. 

Following previous studies~\cite{ESFC,toolace}, we still require data validation and assembly in this process, which involves checking consistency in the tool calling and verifying the accuracy of parameters using the LLM. During data assembly, we will remove duplicates from the merging dataset, and the updated dataset will be used for model training. 

\subsubsection{Model training}

As reinforcement learning with verifiable rewards (RLVR) has become prevalent in training reasoning large language models for reasoning~\cite{deepseek}, we adopt a two-step method to enhance the accuracy and stability of LLM's tool-calling capability, including supervised finetuning (SFT) and reinforcement learning (RL). However, the longer chain-of-thought will introduce significant computational overhead in inference~\cite{overthinking}. Especially in financial QA, some queries aim to obtain up-to-date information via function calling, and a lengthy chain-of-thought will harm the user experience. Therefore, our training needs to strike a balance between accuracy and efficiency.

\begin{figure}[htb]
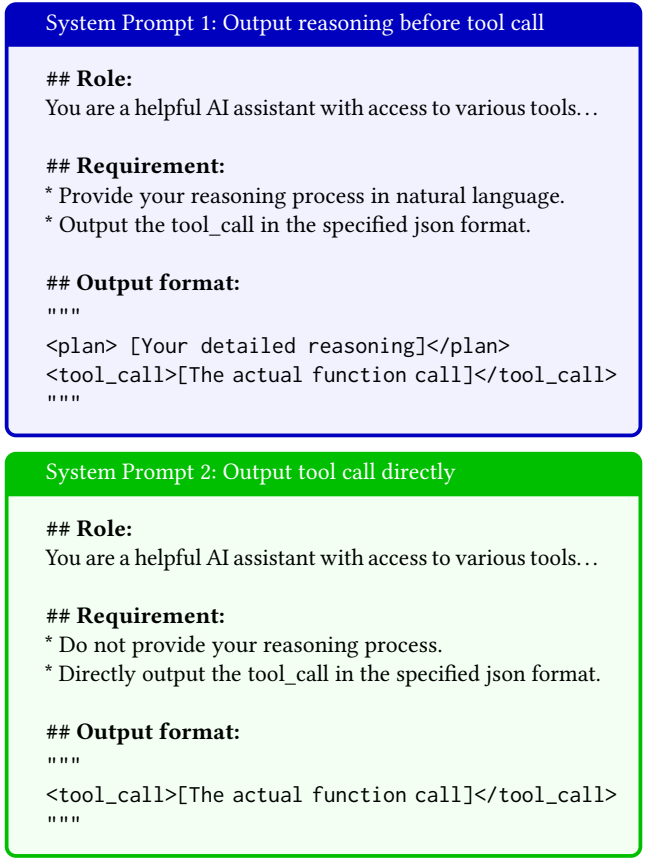

\begin{tcolorbox}[colback=blue!5!white,colframe=blue!75!black,title=System Prompt 1: Output reasoning before tool call]
\textbf{\#\# Role:}\\
You are a helpful AI assistant with access to various tools… \\

\textbf{\#\# Requirement:}

* Provide your reasoning process in natural language. \\
* Output the tool\_call in the specified json format.\\

\textbf{\#\# Output format:}

\begin{verbatim}
"""
<plan> [Your detailed reasoning]</plan>
<tool_call>[The actual function call]</tool_call>
"""
\end{verbatim}

\end{tcolorbox}

\begin{tcolorbox}[colback=green!5!white,colframe=green!75!black,title=System Prompt 2: Output tool call directly]
\textbf{\#\# Role:}\\
You are a helpful AI assistant with access to various tools… \\

\textbf{\#\# Requirement:}

* Do not provide your reasoning process. \\
* Directly output the tool\_call in the specified json format.\\

\textbf{\#\# Output format:}

\begin{verbatim}
"""
<tool_call>[The actual function call]</tool_call>
"""
\end{verbatim}"""
\end{tcolorbox}
\caption{The prompt template for prompt isolation.}
\label{fig:prompt}
\end{figure}

In the SFT step, we will finetune the model with our samples, which provide a good starting point for the next step. The samples consist of two types: reasoning samples $\{r_1,\cdots,r_L,t_1,\cdots,t_m\}$ and direct calling samples $\{t_1,\cdots,t_m\}$. To finetune a model with the mixup of data, we design a prompt isolation as shown in Figure~\ref{fig:prompt}. The system prompt 1 will output reasoning tokens enclosed in $<plan>\cdots</plan>$ before the tool call enclosed in $<tool\_call>,\cdots,</tool\_call>$, and the system prompt 2 will output the tool call directly. During inference, we can use prompt 2 to save the number of tokens for reasoning when necessary.

In the RL step, we adopt a similar rule-based reward formulation that combines format and correctness components. 
The format reward $\mathcal{R}_{format}\in\{0,1\}$ checks whether the model output is consistent with the data format used in the SFT step:

\begin{equation}
\mathcal{R}_{format}=\begin{cases}
1, & \text{if the format is consistent with the input data} \\
0, & \text{otherwise}
\end{cases}
\end{equation}

As to correctness components, we decompose the reward into three components. Suppose the generated tool call list is $a_g=\{t_{g1}(p_{g1}),\cdots,t_{gm}(p_{gm})\}$ and the reference answer is $a_r=\{t_{r1}(p_{r1}),\\ \cdots,t_{rm}(p_{rm})\}$, we define three components as follows:

\begin{itemize}[leftmargin=*]
\item Tool call list retrieval: 
\begin{equation}
F1_t=2 \times \frac{Precision \times Recall}{Precision + Recall},
\nonumber
\end{equation}
where
$Precision=\frac{a_g \cap a_r}{a_g}$, $Recall=\frac{a_g \cap a_r}{a_r}$.

\item Parameter name key retrieval: 
\begin{equation}
F1_p=\frac{1}{m}\Sigma_if1(i), 
\nonumber
\end{equation}
where $f1(i)=2 \times \frac{Precision \times Recall}{Precision + Recall}$, $Precision=\frac{p_{gi} \cap p_{ri}}{p_{gi}}$, $Recall=\frac{p_{gi} \cap p_{ri}}{p_{ri}}$.

\item Parameter value exact matching:
\begin{equation}
EM=\frac{1}{N}\Sigma_{k=1}^N \mathbb{I}(p_{gi}[k]=p_{ri}[k]),
\nonumber
\end{equation}

where $p_{gi}[k]$ and $p_{ri}[k]$ represent the parameter values with respect to the $i$-th parameter.
\end{itemize}

Combining these three values, we get the correctness reward:
\begin{equation}
\mathcal{R}_{correct}=F1_t+F1_p+EM
\end{equation}

Based on the final reward, we can optimize the policy $\pi$ by GRPO~\cite{toolrl}:
\begin{equation}
\begin{aligned}
J_{GRPO}(\theta)=&E_{Q\backsim \mathbb{B}}E_{s\backsim\pi_\theta}[min(\frac{\pi(s_i|Q)}{\pi_{old}(s_i|Q)}A_i(s_i|Q),  \\&clip(\frac{\pi(s_i|Q)}{\pi_{old}(s_i|Q)},1-\epsilon,1+\epsilon)A_i(s_i|Q))
\\ &-\beta KL(\pi \Arrowvert \pi_{ref})]
\end{aligned}
\end{equation}
where $A_i$ is the group normalized advantage. We get a model that can output the reasoning tokens or directly output the function calling results.

\section{Offline experiments}

Due to our pipeline targets in the online setting, we primarily conduct extensive offline experiments to verify two key components of our pipeline: data augmentation and model training.

We first validate our AugFC based on the setting in which it is employed, using the existing benchmark dataset. We also employ our training method on different sizes of base models to verify its superiority. The five main research questions need to be answered as follows:
\begin{itemize}[leftmargin=*]
\item RQ1: How can our AugFC improve the performance on different benchmark datasets with some seed data?

\item RQ2: Do our methods really mitigate the blind spots?

\item RQ3: How does the performance vary with the $\tau_g$ and $\tau_b$ in our AugFC?

\item RQ4: What is the role of some key components in AugFC?

\item RQ5: How does our training method perform compared with existing function calling methods?

\end{itemize}

\subsection{Experimental settings}

\subsubsection{Datasets} 
Considering our single-hop financial QA, we introduce six benchmark datasets to evaluate our method. The \textbf{API-Bank}~\cite{APIbank} comprises two versions, which include 314 tool-use dialogues and 753 API calls, evaluating models' ability to invoke a known API(L-1) or retrieve and call APIs from a candidate list(L2). \textbf{Tool-Alpaca}~\cite{toolalpaca} contains 271 tool-use instances in 50 categories. \textbf{Seal-Tools}~\cite{sealtools} is one of the extensive and recent benchmarks, with 4,076 automatically generated APIs across various life domains. \textbf{Nexus Raven Evaluation}~\cite{nexusraven} consists of 318 test examples across 65 distinct APIs. Lastly, we sample \textbf{xLAM-small} at a ratio of 0.1 from xLAM-60k~\cite{xLAM}, which utilizes over 3,673 APIs across 21 categories from ToolBench~\cite{toolllm}. 

\subsubsection{Metrics}
The tool selection can be evaluated as a multi-class classification task, where each function tool category is treated as a class. Then, a confusion matrix is constructed, where the rows represent the actual tool categories and the columns represent the predicted categories. With the confusion matrix, we can adopt the \textbf{F1 score} to evaluate the model's capability to select tools.

\subsubsection{Base models training}
The language models we adopt in training are the Qwen2.5 series~\cite{qwen}, whose sizes range from 1.5B to 7B and 32B. We first sample $90\%$ xLAM-60k as a seed dataset, and employ our AugFC to generate augmented data. The combined dataset is then used for training our model. 
For comparison with other existing function calling methods, we directly adopt the open-sourced model with different model sizes, including the xLAM~\cite{xLAM} series and the Hammer~\cite{Hammer} series.

\subsection{RQ1: Overall performance of data augmentation}

\begin{table*}[!h]
    \centering
    \begin{tabular}{c|c|c c c c c c |c}
    \toprule
    \textbf{Model Size} & \textbf{Training dataset} & \makecell[c]{\textbf{API-Bank} \\ \textbf{L1}} & \makecell[c]{\textbf{API-Bank} \\ \textbf{L2}}  & \textbf{Tool-Alpaca} & \textbf{Seal-Tools} & \makecell[c]{\textbf{Nexus}\\ \textbf{Raven}} & \textbf{xLAM-small} & \textbf{Avg} \\
    \midrule
    \multirow{4}{*}{Qwen2.5-1.5B} & vanilla & 0.6525 & 0.4293 & 0.4290 & 0.7491 & 0.4988 & 0.5577 & 0.5527 \\
    & src-only & \underline{0.7826} & \underline{0.6533} & \underline{0.6084} & 0.8865 & 0.5348 & 0.5934 & \underline{0.6765} \\
    & aug-only & 0.6973 & 0.4992 & 0.4242 & \underline{0.8551} & \underline{0.7500} & \underline{0.6857} & 0.6519 \\
    & src+aug & \textbf{0.7226} & \textbf{0.6648} & \textbf{0.6233} & \textbf{0.8898} & \textbf{0.7840} & \textbf{0.7288} & \textbf{0.7256} \\
    \midrule
    \multirow{4}{*}{Qwen2.5-7B} & vanilla & 0.6985 & 0.6036 & 0.6158 & 0.8083 & 0.7283 & 0.5914 & 0.6743 \\
    & src-only & \textbf{0.8295} & \textbf{0.6529} & \textbf{0.6745} & \underline{0.9426} & \underline{0.7892} & 0.7428 & \underline{0.7719} \\
    & aug-only & 0.7776 & 0.5824 & 0.5808 & 0.9116 & 0.6238 & \underline{0.8038} & 0.7133 \\
    & src+aug & \underline{0.8002} & \underline{0.6440} & \underline{0.6667} & \textbf{0.9492} & \textbf{0.8901} & \textbf{0.8165} & \textbf{0.7856} \\
    \midrule
    \multirow{4}{*}{Qwen2.5-32B} & vanilla & 0.7988 &\underline{0.6304} & 0.6448 & 0.9339 & 0.8266 & 0.7135 & 0.7580 \\
    & src-only & \underline{0.8325} & 0.5433 & \textbf{0.6792} & \underline{0.9485} & \textbf{0.8807} & 0.7736 & \underline{0.7763} \\
    & aug-only & 0.7771 & 0.5357 & 0.6308 & 0.9129 & 0.8624 & \underline{0.8282} & 0.7579 \\
    & src+aug & \textbf{0.8416} & \textbf{0.6501} & \underline{0.6775} & \textbf{0.9494} & \underline{0.8682} & \textbf{0.8479} & \textbf{0.8058} \\
    \bottomrule
    \end{tabular}
    \caption{The overall performance to validate our AugFC. The vanilla denotes the base model without any training, the src-only denotes we only use the seed data, aug-only means the augmented data is used, and src+aug represents the combined dataset. All results are measured using the F1 score. The best results are bold, and the second-best results are underlined.}
    \label{tab:qwen2.5-results}
\end{table*}

To evaluate the data augmentation, we compared the performance of models trained on different datasets. Specifically, the vanilla represents the base model, relying on the base model's capability to invoke a function. The src-only denotes model training based on our sampled xLAM-60k, while the aug-only uses augmented data based on the src-only model with our AugFC. src+aug is our combined dataset, which is consistent with our pipeline. The overall performance is given in Table~\ref{tab:qwen2.5-results}. 

Based on the results, we make the following observations. First, the base model's function calling capability fails to meet the accuracy requirements. Notably, the small model can hardly achieve comparable performance with trained models, which justifies the need to investigate this problem. Second, the src-only data performs better than the aug-only data, indicating that the augmented data cannot achieve comparable performance in the absence of the original data and sometimes degrades its performance. Ultimately, our combined dataset, which encompasses both source and augmented data, yields the best performance on average. This further validates the effectiveness of our AugFC.

\subsection{RQ2: The number of blind spots}

\begin{figure}[!htb]
    \centering
    \includegraphics[width=\linewidth]{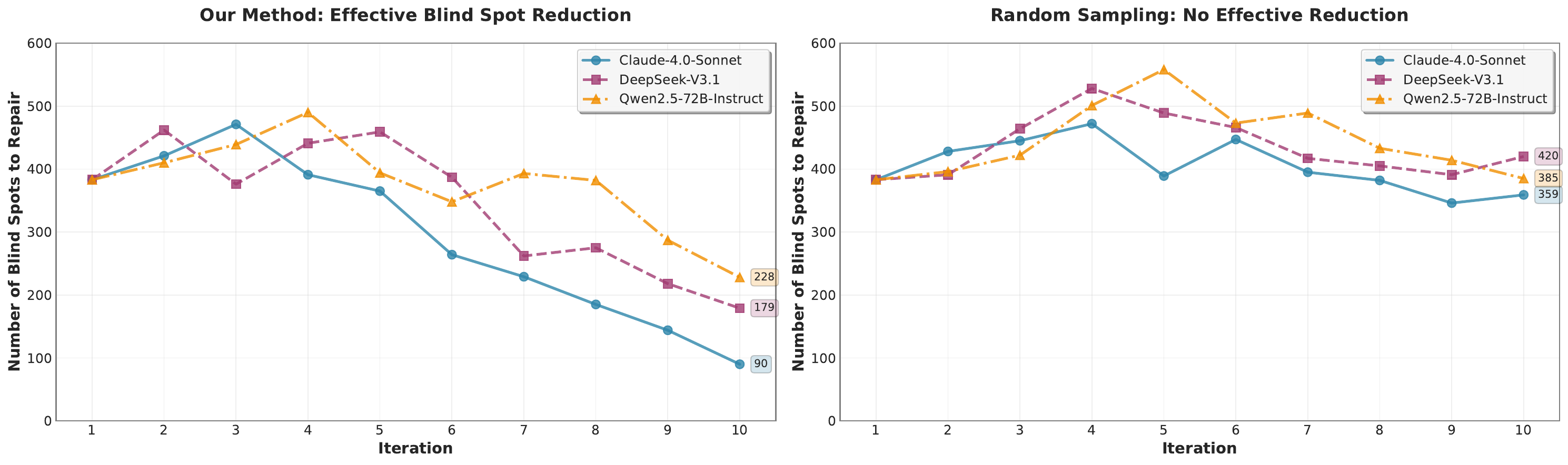}
    \caption{The number of blind spots to repair with different LLMs.}
    \label{fig:blind}
\end{figure}

The key concept in AugFC is the existence of blind spots. We introduce a powerful language model to generate related data and repair blind spots. Hence, we will answer whether the number of blind spots decreases with each iteration. We utilize three different LLMs as generated LLMs, including both commercial and open-source models, as shown in Figure~\ref{fig:blind}, including Claude-4~\footnote{https://claude.ai/}, Deepseek-v3.1~\footnote{https://chat.deepseek.com/}, and Qwen2.5-72B-instruct~\footnote{https://chat.qwen.ai/}. 

Meanwhile, we also use random sampling as a comparison, which draws data directly from users' online queries. It is evident that our method significantly reduces blind spots compared to random sampling across three main LLMs. Moreover, random sampling may introduce new blind spots due to its random nature.

\subsection{RQ3: The effect of hyperparameters.}

\begin{figure}[!htb]
    \centering
    \includegraphics[width=1\linewidth]{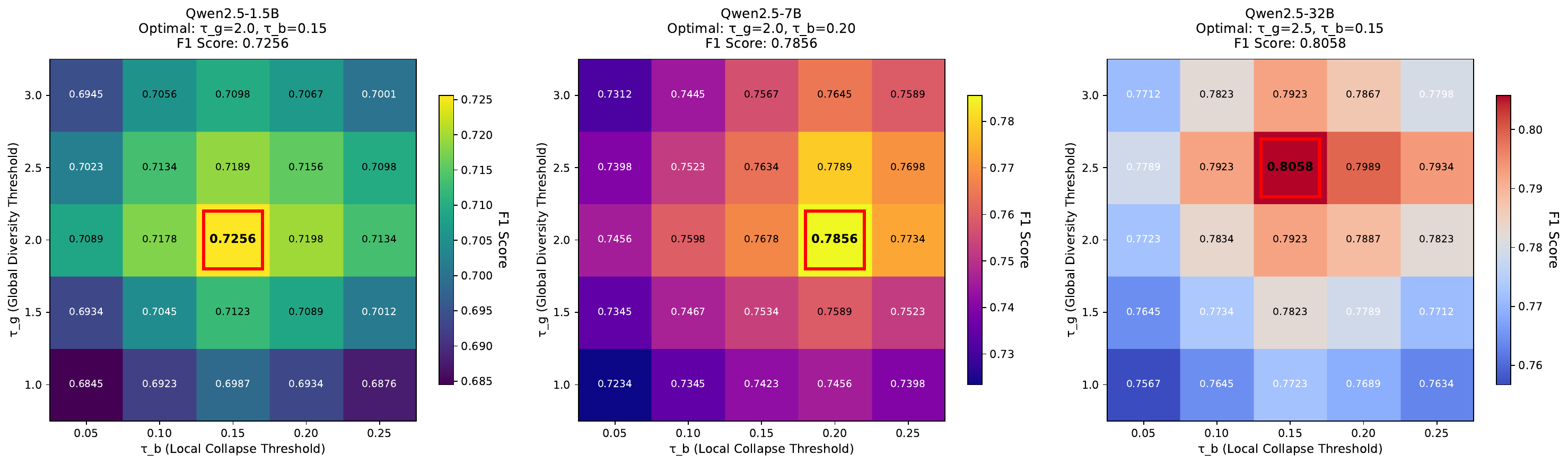}
    \caption{The heatmap illustrating how $\tau_g$ and $\tau_b$ affects the performance resepectively.}
    \label{fig:heatmap}
\end{figure}

Referring to Definition~\ref {def:blind}, two key hyperparameters $\tau_g$ and $\tau_b$ determine the blind spot together. Therefore, we need to investigate how these two hyperparameters affect our pipeline. We do a grid search on two hyperparameters, where $\tau_g \in \{1.0,1.5,2.0,2.5,3.0\}$ and $\tau_b \in \{0.05,0.1,0.15,0.2,0.25\}$ for each size of base model. The heat maps are illustrated in Figure~\ref {fig:heatmap}. 

Notice that the $\tau_g$ is not the largest one that leads to better performance. The reason is that the large one will filter out some queries that do not need to be augmented to conquer blind spots with data augmentation. Meanwhile, $\tau_b$ is not the smallest one that leads to better performance, which indicates the difficulty in repairing blind spots when the parameter distribution collapses too much.

\subsection{RQ4: The ablation study of AugFC}

To verify the effectiveness of the components in AugFC, we conduct an ablation study of AugFC. The first one is \textit{w/o blind spots}, which uses random sampling instead of blind spots detection. The second is \textit{w/o designed prompt}, which uses a plain prompt without distribution information. Lastly,\textit{w/o multi-round} indicates that we only use a single round to generate the augmented data. The results averaged on the benchmark datasets are illustrated in Table~\ref{tab:ablation}.

From the results, we conclude that the \textit{designed prompt} has the most significant impact on performance, with reductions of $7.4\%$, $6.1\%$, and $4.2\%$ on the three-sized models. The \textit{designed prompt} directly determines the quality of the generated samples. The component of \textit{blind spots} also significantly affects performance, because it provides information on which parameters need to be augmented. The absence of \textit{multi-round} also degrades the performance, indicating that multi-round is also necessary. The overall ablation further verifies the effectiveness of our AugFC.

\begin{table}[!h]
    \centering
    \scalebox{0.9}{
    \begin{tabular}{c|c c c c }
    \toprule
    Model & AugFC & \makecell[c]{\textit{w/o} \\ \textit{blind spots}} & \makecell[c]{\textit{w/o} \\ \textit{designed prompt}} & \makecell[c]{\textit{w/o}\\\textit{multi-round}}\\
     \midrule
     Qwen2.5-1.5B & 0.726 & 0.689 & 0.672 & 0.695 \\
     \midrule
     Qwen2.5-7B & 0.786 & 0.742 & 0.738 & 0.752 \\
     \midrule
     Qwen2.5-32B & 0.806 & 0.765 & 0.772 & 0.780 \\
     \bottomrule
    \end{tabular}
    }
    \caption{The ablation study of AugFC. The results are the averages of the six benchmark datasets.}
    \label{tab:ablation}
\end{table}

\subsection{RQ5: The effectiveness of two-step method}

Lastly, we conduct experiments to verify our proposed two-step method. We compare Hammer, xLAM, and Qwen2.5 with our model on six benchmark datasets. xLAM uses the SFT approach, further aligning model checkpoints with the DP method. Hammer only adopts the finetuning step. Notice that the maximum size of the hammer series is 7B. Both xLAM-1.3B, xLAM-7B, and the Hammer series are trained on the xLAM-60k dataset, while the Qwen2.5 models serve as the base models. The results are demonstrated in Table~\ref{tab:func_param}.

\begin{table*}[!h]
    \centering
    \begin{tabular}{ccccccccc}
    \toprule
    \textbf{Model Size} & \textbf{Model Type} & \makecell[c]{\textbf{API-Bank} \\ \textbf{L1}} & \makecell[c]{\textbf{API-Bank} \\ \textbf{L2}}  & \textbf{Tool-Alpaca} & \textbf{Seal-Tools} & \makecell[c]{\textbf{Nexus}\\ \textbf{Raven}} & \textbf{xLAM-small} & \textbf{Avg}  \\
    \midrule
    \multirow{4}{*}{\textbf{1.5B}} 
        & Qwen2.5-1.5B & 0.6525 & 0.4293 & 0.4290 & 0.7491 & 0.4988 & \underline{0.5577} & 0.5527 \\
        & xLAM-1.3B-fc & \textbf{0.8370} & \textbf{0.6432} & 0.5058 & 0.8043 & 0.5480 & 0.5368 & \underline{0.6459} \\
        & Hammer-1.5B & 0.7230 & 0.5971 & \underline{0.5348} & \underline{0.8865} & \underline{0.5688} & 0.5192 & 0.6382 \\
        & ours & \underline{0.7226} & \underline{0.6048} & \textbf{0.6233} & \textbf{0.8898} & \textbf{0.7840} & \textbf{0.7288} & \textbf{0.7256} \\
    \midrule
    \multirow{4}{*}{\textbf{7B}} 
        & Qwen2.5-7B & 0.6985 & 0.6036 & 0.6158 & 0.8083 & 0.7283 & 0.5914 & 0.6743 \\
        & xLAM-7B-fc & \underline{0.8069} & 0.6424 & 0.5896 & 0.7687 & 0.5409 & 0.6378 & 0.6644 \\
        & Hammer-7B & \textbf{0.8311} & \textbf{0.6598} & \underline{0.6250} & \underline{0.8987} &\underline{0.7464} & \underline{0.6535} & \underline{0.7358} \\
        & ours & 0.8002 & \underline{0.6440} & \textbf{0.6667} & \textbf{0.9492} & \textbf{0.8901} & \textbf{0.8165} & \textbf{0.7856} \\
    \midrule
    \multirow{4}{*}{\textbf{32B}} 
        & Qwen2.5-32B & 0.7988 & \underline{0.6304} & 0.6448 & \underline{0.9339} & \underline{0.8266} & \underline{0.7135} & 0.7580 \\
        & xLAM-2-32B-fc & \underline{0.8270} & 0.6000 & \underline{0.6597} & 0.9049 & 0.8573 & 0.6985 & \underline{0.7652} \\
        & ours & \textbf{0.8416} & \textbf{0.6501} & \textbf{0.6775} & \textbf{0.9494} & \textbf{0.8682} & \textbf{0.8479} & \textbf{0.8058} \\
    \bottomrule
    \end{tabular}
    \caption{The comparison of our two-step training model with existing models. All results are measured using the F1 score. The best results are bold, and the second-best results are underlined.}
    \label{tab:func_param}
\end{table*}

The base model performs the worst, which is in accordance with the conclusions based on Table~\ref{tab:qwen2.5-results}. It is then clear that our training method outperforms the other two across all model sizes, indicating that the two-step method is superior to other finetuning methods. 

\section{Online experiments}

Our data-driven pipeline is primarily centered on the online setting in FiT, Tencent. We also conduct various experiments on our online financial QA systems, which have been integrated into Yuanbao. 

\subsection{Online system}

\begin{figure*}[!htb]
    \centering
    \includegraphics[width=0.9\linewidth]{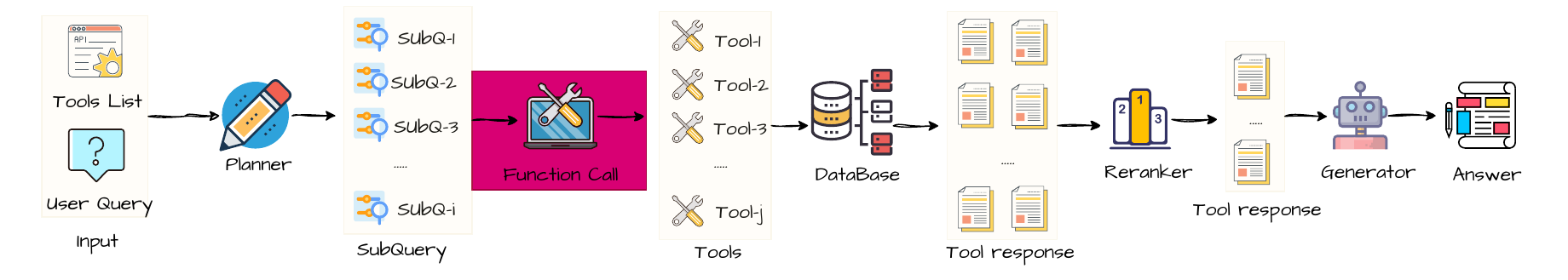}
    \caption{The overview illustration of the online financial QA system. Our pipeline is marked as red in the whole system.}
    \label{fig:system}
\end{figure*}

We first briefly present our online financial QA system in Figure~\ref{fig:system}. The system consists of four components: planner, function call, reranker, and generator. The planner will provide the subqueries based on the tool list and the user query. The function call is where our pipeline works. After our pipeline generates the tool response, the reranker will rerank the inputs according to relevance and other constraints. A generator will output the answer. Obviously, the function call component mainly determines the quality of the answer by involving tools. 

\subsection{Online results}

\begin{figure}
    \centering
    \includegraphics[width=1\linewidth]{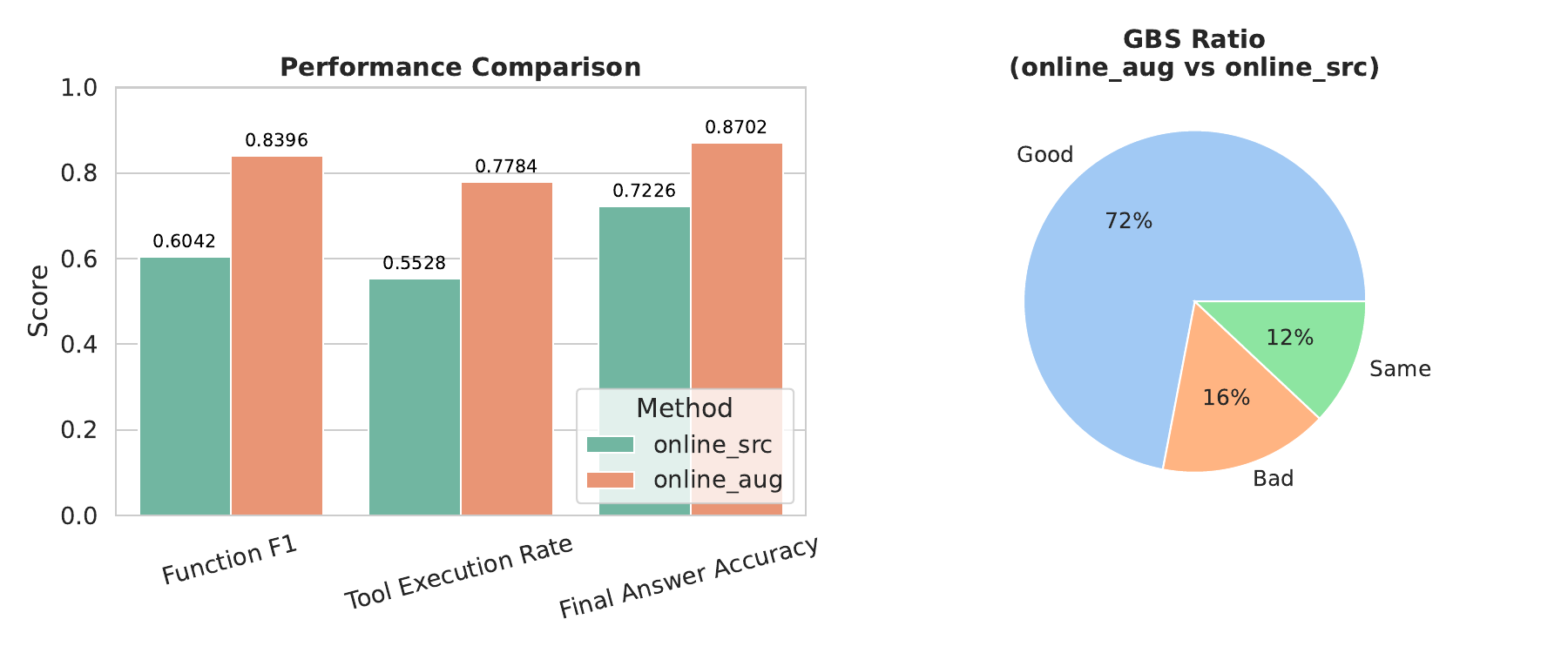}
    \caption{The online results measured by four metrics.}
    \label{fig:online}
\end{figure}

In the online setting, we deploy two pipelines in the function call. One is our data-driven pipeline, and another one is without our pipeline. Specifically, we have an annotated dataset named \textit{online\_src} at the beginning. After our pipeline, the augmented dataset \textit{online\_aug} is then obtained. The base model only trains the model with \textit{online\_src} using SFT, whereas our pipeline provides a model with \textit{online\_aug}.

The online metrics encompass both automated and manual ones. The automated ones consist of \textbf{F1 score} and \textbf{Tool Execution Rate}. The F1 score is consistent with our offline experimental setting, indicating accuracy. The tool execution rate refers to the success rate of tool execution in a real-world environment, indicating the reliability of the pipeline. The manual ones include \textit{Final answer accuracy} and \textit{GBS ratio}. The final answer accuracy is checked by determining whether the final answer is aligned with the reference answer, ignoring the intermediate output. The GBS ratio is Good: Bad: Same ratio, which is an evaluation metric used to assess the impact of changes in complex systems manually. 

A query will traverse the entire system, whether or not our pipeline is involved, and the online metrics will measure both results. We evaluate 350 end-to-end query-answer pairs, which involve nearly 2,000 pairs of query and tool-call list. The results are demonstrated in Figure~\ref{fig:online}. For the automated metrics, the F1 score improves by $39.1\%$, and the tool execution rate gains $40.7\%$ improvements. For manual metrics, final answer accuracy gains $20.3\%$ improvements, and the GBS ratio has $72\%$ good results compared with $16\%$ bad ones.
It can be concluded that our pipeline is superior to the baseline and meets the requirements for full deployment on the system. 

Moreover, we randomly sampled 500 instances (10\% of the augmented dataset) and assigned them to two experts for independent review. 90\% of the augmented data are directly usable, 6\% contain minor issues that can be easily corrected, and only 4\% exhibit severe errors requiring discarding.

\section{Conclusion}
In this paper, we present a data-driven pipeline to improve function calling in LLM for our online financial QA system. The pipeline collects online user queries to exploit the toolset and proposes a data augmentation method, AugFC, to explore potential queries, thereby addressing the blind spots in the query space. A two-step training method is then introduced to enhance the capability to call functions. 
We hope our work can provide practitioners with experience of deploying LLMs in real-world scenarios, and the data-driven approach will offer a new perspective on LLM applications. In the future, we will extend our pipeline to include multiple tool dependencies or cross-module call scenarios to further verify effectiveness.

\section{Acknowledgement}
We thank the support of the Shenzhen Technology University School-level
(No.20251061020002), Scientific Research Capacity Enhancement Program for Key Construction Disciplines in Guangdong Province (No.2024ZDJS063), the National Natural Science Foundation of China(NSFC) (No. 62506238).
\clearpage
\bibliographystyle{ACM-Reference-Format}
\bibliography{reference}
\clearpage
\appendix
\section{Prompt list used in pipeline}

\begin{tcolorbox}[
    colback=purple!5!white,
    colframe=purple!75!black,
    title=\textbf{Prompt: Tool Call Consistency Checker},
    fonttitle=\bfseries,
    breakable,
    enhanced,
    width=\textwidth,
    left=2mm,
    right=2mm,
    top=2mm,
    bottom=2mm,
    label=fig:algebra-misconception
]
\small\ttfamily
\textbf{\#\# Role:}
You are a Tool Call Consistency Checker. \\

\textbf{\#\# You will receive:} \\
- Generated user query: \{\{query\}\}

- Tool definition: \{\{tools\}\}

- Generated tool\_call JSON: \{\{tool\_call\}\} \\

\textbf{\#\# Your task:}\\
1. Carefully read the generated user query and understand the \textbf{intended action}. \\
2. Review the tool definition to understand each parameter's \textbf{meaning and constraints}. \\
3. Check the parameter values in tool\_call: \\
   \hspace*{2em}- Do they match the intent and details in the user query? \\
   \hspace*{2em}- Are they internally consistent (no contradictions between parameters)? \\
   \hspace*{2em}- Do they comply with the tool definition (value types, required fields, allowed ranges)? \\
4. Decide if the tool\_call is logically correct: \\
   \hspace*{2em}- If all parameters reflect the query correctly and satisfy the tool's definition, return \textbf{"Consistent"}. \\
   \hspace*{2em}- If there is any mismatch or logical conflict, return \textbf{"Inconsistent"}. \\

\textbf{\#\# Output format requirements:} \\
- Return a (\textbf{JSON list}) containing exactly one object:
\begin{verbatim}
[{
    "analysis": "...your reasoning here...",
    "result": "Consistent" or "Inconsistent"
}] 
\end{verbatim}
\vspace{\baselineskip}

\textbf{\#\# Rules:} \\
\hspace*{2em}- Do not output anything outside the JSON list. \\
\hspace*{2em}- Be strict - even small inconsistencies should be marked \textbf{"Inconsistent"}.

\end{tcolorbox}

\onecolumn
\begin{tcolorbox}[
    colback=orange!5!white,
    colframe=orange!75!black,
    title=\textbf{Prompt: Few-Shot Tool Call JSON Generator},
    fonttitle=\bfseries,
    breakable,
    enhanced,
    width=\textwidth,
    left=2mm,
    right=2mm,
    top=2mm,
    bottom=2mm,
    label=fig:algebra-misconception
]
\small\ttfamily
\textbf{\#\# Role:} \\
You are an expert assistant capable of accurately selecting and calling functions (tools) to answer questions. \\

\textbf{\#\# Input:} \\
1. A set of FIVE few-shot examples, each containing: \\
   \hspace*{2em}- A user query\\
   \hspace*{2em}- A toolset with tool names, descriptions, and parameters\\
   \hspace*{2em}- The correct tool calls for that query in strict JSON list format\\
2. The CURRENT user query we need to process\\
3. The CURRENT toolset specification\\

\textbf{\#\# Your task:}\\
\hspace*{2em}- Carefully study the five examples to understand how queries are mapped to tool calls.\\
\hspace*{2em}- For the CURRENT query, use ONLY the tools provided in the CURRENT toolset, along with their descriptions, to determine the exact functions to call and the correct values of their parameters.\\
\hspace*{2em}- Parameter values MUST be derived accurately from the query context or the tool definitions.\\
\hspace*{2em}- If no tool is required, output an empty list: \texttt{[]}.\\

\textbf{\#\# Output requirements:} \\
\hspace*{2em}- You MUST follow the exact JSON list format below.\\
\hspace*{2em}- DO NOT include any extra explanations, comments, or text outside the JSON.\\
\hspace*{2em}- Ensure parameter types are correct (string, integer, float, etc.).\\

\textbf{\#\# Expected JSON format:}
\begin{verbatim}
[{
    "name": "func_name1", "arguments": {"argument1": "value1", "argument2": "value2"}},
    ... (more tool calls as required)
}]
\end{verbatim}
\vspace{\baselineskip}

\textbf{\#\#\# Few-shot Examples:}\\
\{\{FEW\_SHOT\_EXAMPLES\}\}\\

\textbf{\#\#\# Current Query:}\\
\{\{CURRENT\_QUERY\}\}\\

\textbf{\#\#\# Current Toolset:}\\
\{\{CURRENT\_TOOLSET\}\}\\

\textbf{Please output the JSON list strictly according to the specifications above.}

\end{tcolorbox}

\onecolumn
\begin{tcolorbox}[
    colback=blue!5!white,
    colframe=blue!75!black,
    title=\textbf{Prompt: Multi-Round Distribution-Aware Counterfactual Generation},
    fonttitle=\bfseries,
    breakable,
    enhanced,
    width=\textwidth,
    left=2mm,
    right=2mm,
    top=2mm,
    bottom=2mm,
    label=fig:algebra-misconception
]
\small\ttfamily
\textbf{\# Role:} \\
A specialist in mitigating data bias through multi-round distribution-aware counterfactual generation.\\

\textbf{\# Multi-Round Generation Context (Step \{step\}):}\\
We are in a multi-round generation process to mitigate distribution collapse in parameter "{tool\_param}".\\

\textbf{\#\# Initial State (Before Generation):}\\
\hspace*{2em}- Global Entropy: \{initial\_state['global\_entropy']:.4f\}\\
\hspace*{2em}- Local Entropy: \{initial\_state['local\_entropy']:.4f\}\\
\hspace*{2em}- Entropy Ratio: \{initial\_state['entropy\_ratio']:.4f\}\\
\textbf{\#\# Current State (After \{step-1\} rounds):}\\
\hspace*{2em}- Global Entropy: \{current\_state['global\_entropy']:.4f\} (change: \{current\_state['global\_entropy'] - initial\_state['global\_entropy']:+.4f\})\\
\hspace*{2em}- Local Entropy: \{current\_state['local\_entropy']:.4f\} (change: \{current\_state['local\_entropy'] - initial\_state['local\_entropy']:+.4f\})\\
\hspace*{2em}- Entropy Ratio: \{current\_state['entropy\_ratio']:.4f\} (change: {current\_state['entropy\_ratio'] - initial\_state['entropy\_ratio']:+.4f})\\
\hspace*{2em}- Target: Increase entropy ratio to $\geqslant$ {blind\_entropy\_ration\_threshold} \\
\hspace*{2em}- History: \{history\_desc\} \\

\textbf{\# Parameter Value Distributions:}\\
\textbf{\#\# Initial Distributions:}\\
GLobal: \{initial\_global\_dist\_desc\}, Local: \{initial\_local\_dist\_desc\}\\
\textbf{\#\# Current Distributions:}\\
Glocal: \{current\_global\_dist\_desc\}, Local: \{current\_local\_dist\_desc\}\\

\textbf{\# Instructions}\\
\hspace*{2em}* Contains an instruction for tool usage: """\{instruction\}"""\\
\textbf{\# History}\\
\hspace*{2em}* Contains the user's historical conversation information: """\{input\_text\}"""\\
\textbf{\# Original Example:}\\
\hspace*{2em}- Query: "\{user\_query\}", Tool Call: \{tool\_call\}\\

\textbf{\# Stable Parameter Context:}\\
To prevent creating new distribution collapses, the values for the following parameters in the `new\_tool\_call` MUST remain consistent with their existing distributions.\\
Your primary goal is to fix `{tool\_param}`, but a CRITICAL secondary goal is to NOT disrupt these other parameters.\\

\textbf{\# Task for Round \{step\}:}\\
Based on the multi-round generation progress, generate \textbf{NEW} data points to further increase local parameter diversity:\\
1. **Learn from previous rounds**: Analyze what values were generated and their impact.\\
2. **Focus on current gaps**: Target parameter values that are still underrepresented in local distribution.\\
3. **Avoid redundancy**: Don't generate values that were already created in previous rounds.\\
4. **Strategic selection**: Choose values that will maximally increase the entropy ratio.\\
5. **Maintain coherence**: Ensure semantic consistency within the cluster context.\\
6. **Diversify query expressions**: Generate queries with varied linguistic forms but similar semantics, avoiding mere entity substitution. \\
7. **Preserve tool\_call logical coherence**: All parameter values in the generated tool\_call must maintain logical consistency. \\

\textbf{\# JSON Output Format:}
\begin{verbatim}    
[{
    "new_query": "...",
    "new_value_for_{tool_param.split('.')[-1]}": "...",
    "new_tool_call": "...",
    "step_rationale": "Strategic explanation for Round {step}"
}]
\end{verbatim}
\vspace{\baselineskip}


\end{tcolorbox}

\end{document}